\documentstyle[12pt]{article}
\newcommand{\be}{\begin{equation}}\newcommand{\ee}{\end{equation}}
\begin{document}
\begin{titlepage}
\vspace*{-1cm}
\noindent
\hfill{CERN-TH.6533/92}
\\
\phantom{bla}
\hfill{RU-92-11-B}
\\
\phantom{bla}
\hfill{hepth@xxx/9207089}
\\
\vskip 2.5cm
\begin{center}
{\Large\bf From 2D conformal to 4D self-dual theories:
quaternionic analyticity }
\end{center}
\vskip 1.5cm
\begin{center}
{\large M.~Evans}\footnote{Rockefeller University, New York, NY
10021-6399, USA. Internet: evans@physics.rockefeller.edu.
Supported in part by the US Department of Energy, grant
no.~DOE-ACO2-87ER40325, TASK B.} \\
\vskip .3cm

{\large F.~G\"ursey}\footnote{Department of Physics, Yale University,
New Haven, CT 06511, deceased.}   \\
\vskip .3cm
and \\
\vskip .3cm
{\large V.~Ogievetsky}\footnote{On leave from Laboratory of Theoretical
Physics, Joint Institute for Nuclear Research, Dubna,
Russia}$^,$\footnote{Department de Physique Th\'eorique, Universit\'e
de Gen\`eve, Gen\`eve, Switzerland. Partly supported by
Swiss National Foundation.}       \\
\vskip .3cm
Theory Division, CERN, \\
Geneva, Switzerland \\
\vskip 1cm
\end{center}
\begin{abstract}
\noindent
It is shown that self-dual theories generalize to four dimensions
both the conformal and analytic aspects of two-dimensional conformal
field theories.
In the harmonic space language there appear several
ways to extend complex analyticity (natural in two dimensions)
to quaternionic analyticity (natural in four dimensions).
To be analytic, conformal transformations should
be realized on $CP^3$, which appears as the coset of the
complexified conformal group modulo its maximal
parabolic subgroup. In this language one visualizes
the twistor correspondence of Penrose and Ward and  consistently
formulates the analyticity of Fueter.
\end{abstract}
\vfill{
CERN-TH.6533/92
\newline
\noindent
{June 1992}}
\end{titlepage}
\setcounter{footnote}{4}
\setcounter{section}{0}

\section*{Feza G\"ursey}
Feza G\"ursey, a fine human being and outstanding physicist,
passed away on April 13, 1992. He is a coauthor of the present paper,
which is one of a series of his works devoted to quaternionic
aspects of four-dimensional field theories, a field in which he
was a pioneer.
Feza enthusiastically participated in the writing of this paper, even
as he fought the disease to which he finally succumbed.
Sadly, he did not live long enough to approve the paper's final version,
and so bears no responsibility for whatever shortcomings it
may possess. It was a great joy and privilege to work with
Feza, and to benefit from his fertile mind and keen intelligence.
The experience of working with him and the wonderful
personality of Feza G\"ursey will abide forever in the memories of the
two other authors.

\section{Introduction}

\subsection{From 2D complex to 4D quaternionic analyticity}
The two {\it real} coordinates of two-dimensional Euclidean space
are quite naturally combined into {\it a single complex number}
\be
x^\mu=\{x^1, x^2\}  \hspace{10mm} \longrightarrow \hspace{10mm}
z=x^1 + i x^2
\ee
As is well known, the most general conformal coordinate transformation
in two ({\it
and only in two}) dimensions is {\it analytic} in this complex coordinate.
\be
z'=f(z), \hskip15mm  \bar z' = \bar f(\bar z).  \label{0}
\ee
Owing to the Cauchy-Riemann condition,  its d'Alembertian vanishes
\be
{\partial\over\partial {\bar z}} f(z)= 0 \hskip5mm \longrightarrow \hskip5mm
\Box f(z)={\partial\over\partial z}{\partial\over\partial{\bar z}}f(z)=0.
\label{1}
\ee

In any higher dimension, conformal transformations depend on a finite
number of parameters, and the d'Alembertian of
 infinitesimal conformal boosts {\it does not vanish}.

In four dimensions, coordinates are well known to be unified into
a quaternion as naturally as, in two dimensions, they are unified into a
complex number. Specifically, in the spinor formalism we have
$$
x^m=\{x^0, x^1, x^2, x^3\} \longrightarrow  z= x^{\alpha \dot\alpha}=
\left(\begin{array}{cc} x^0 - ix^3 & -ix^1 - x^2 \\
                       -ix^1 + x^2 & x^0 + ix^3 \end{array}\right) =    $$
\be
x^0 I -i\sigma_a x^a = x^0 + e_a x^a         \label{2}
\ee
and the Pauli matrices represent the algebra of the quaternionic  units,
$e_a=-i \sigma_a$
\be
e_a e_b =- \delta_{ab} + \epsilon_{abc} e_c.     \label{3}
\ee

The analytic transformations (\ref{0}) are fundamental to
2D-conformal field theories.
It is natural to wonder whether there exist
4D theories in which some form of quaternionic
analyticity would play a corresponding r\^ole \cite{oldgu}, \cite{gur1}.
However, the notion of quaternionic analyticity proves to be rather
delicate, and we shall see that there are several potential forms,
only some of which will prove interesting (see, e.g. \cite{sud}).

\subsection{Difficulties with quaternionic analyticity}
A straightforward extension of the Cauchy-Riemann condition would be
\be
{\partial\over\partial {\bar z}} f=
\frac{1}{2} \left ({\partial\over\partial{x^0}}+
\frac{1}{3} e_a {\partial\over\partial {x^a}}\right ) f=0      \label{4}
\ee
where ${\partial\over\partial{\bar z}}$ has been defined in such a way that
${\partial\over\partial{\bar z}} z = 0 $ and
${\partial\over\partial{\bar z}} \bar z = 1 $.
It is well known however (see e.g., \cite{sud}) that the only solution
to equation (6) in the form of a power series in $z$ is
$f= a + zb $, with constant
quaternions $a$ and $b$, owing to the noncommutativity of quaternions.
Even ${\partial\over\partial{\bar z}} z^2 = \frac{1}{3} (z - \bar z )$.

Heretofore, {\em Fueter quaternion analyticity} \cite{fuet}, \cite{gurs}
\cite{gj} was the only successful attempt
to find something less restrictive than (\ref{4}). An analytic
function of a quaternion, $z$, is defined by a Weierstrass-like series
\be
f(z) = \sum a_n z^n,          \label{F}
\ee
where the coefficients $a_n$ are real or complex numbers
(or quaternions, but standing only to the one side of $z^n$, say,
left as in (\ref{F})). Such a function can be shown to obey
some Cauchy-Riemann-
like condition, but of {\it the third order} in derivatives instead of the
first. The equation $\Box f(z)=0 $ does not hold in general;
however the equation $ \Box^2 f(z) =0 $ is preserved.
It was emphasized in \cite{gj}, that the above definition may
hold in some $SO(4)$ frames of reference, but not in others\footnote{
Indeed, four-dimensional rotations
$SO(4)\simeq SU(2)_L\times SU(2)_R$ are known to have the
quaternionic form \cite{oldgu}, \cite{gj} $
z' = m z \bar n, \hskip5mm  m \bar m = n \bar n = 1, $
where $m\in SU(2)_L$ and $n\in SU(2)_R$ are unit quaternions
representing these groups.
There are problems already with the $z^2$ term.
It is transformed into $mz\bar n m z \bar n$,  which
cannot be expressed in the initial analytic form
owing to the non-commutativity of quaternions.}.

Despite these difficulties, in
the self-dual theories \cite{gib} and in the $N=2$
supersymmetric theories \cite{alv}, \cite{bag},
there arise manifolds of a quaternionic character,
namely quaternionic-K\"ahler and hyperK\"ahler manifolds. Indeed,
the problems associated with quaternionic analyticity are very
reminiscent of the difficulties involved in finding a non-trivial
notion of quaternionic geometry. It is tempting to speculate that
the paucity of solutions to equation (6) is the analytic manifestation
of the fact that only flat metrics are hyperk\"ahler with respect to
integrable almost quaternionic structures.

Therefore the fact that interesting quaternionic geometries {\it do\/}
exist suggests that it should be possible to find a useful notion of
quaternionic
analyticity. It is realized within the harmonic \cite{gal}
(a kind of twistor \cite{pen}, \cite{ward}, \cite{wardw}, \cite{penr},
\cite{bas}, etc.) space approach which has
proven to be effective in theories possessing quaternionic
structures \cite{gal}, \cite{galp}, \cite{nhi},
\cite{og}, \cite{sok}, \cite{baggal},
etc., as well as for properly improving \cite{gurs}, \cite{gj} the
above Fueter definition (see also Section 5).
\subsection{Plan and results}
The aim of the present note
is to show that the harmonic space approach opens new horizons in a search
for useful definitions of quaternionic analyticity, including those with
Cauchy-Riemann conditions
of the first order in derivatives, and with analytic functions with
vanishing d'Alembertians. This approach helps
to achieve these goals.

There are several ways to implement a harmonic version of quaternionic
analyticity.
One way leads just to the self-dual Yang-Mills and Einstein theories,
which thus appear as 4D counterparts to  2D conformal field
theories, in the sense that both are
analytic in the dimensionally appropriate sense. We shall deal in this
paper with a presentation of the self-dual Yang-Mills theories in
harmonic space and with their analytic structure there.

Another way leads to a ``covariantization" of the Fueter definition
\cite{gurs} and
the corresponding coset space will be discussed also.

The tempting problem of finding a four-dimensional counterpart to the two-
dimensional conformal field theories has been attacked by a number of
authors, see in particular recent
papers, such as \cite{sin},  \cite{vafa}. The 4D
self-dual gauge and gravity theories were considered very promising
candidates. It must also be mentioned that intimate connections of these
theories to various one- and two-dimensional integrable systems were
discussed more than ten years ago already, e.g. \cite{bel}, \cite{cor},
\cite{lez}.
A suggestion was even made \cite{wardint}, \cite{hit} that all
integrable systems might be deduced by dimensional reduction from the
4D self-dual theories, inheriting their remarkable properties. A number
of recent papers (\cite{mas}, \cite{new}, \cite{park}, \cite{bak},
\cite{chau}, \cite{tak}, etc.) provide strong support for this suggestion,
revealing also the importance of both signatures, (4,0) and (2,2).

The harmonic space approach allows a systematic study of the self-dual
theories and their symmetries based on their quaternionic analyticity
(in the harmonic sense), which replaces the standard
complex one of the 2D conformal theories. The conformal invariance of
self-duality also plays an essential role: a) It puts
space and harmonic coordinates on an equal footing,
b) The requirement that conformal transformations must be analytic
leads to the remarkable phenomenon of {\it complexification} --
they have to be realized as a real $Spin(5,1)$ group acting
on a 5-dimensional compact coset of its complexification, $Spin(6,C)
\sim SL(4,C)$.
This coset is just
$CP^3$. This manifold is complex with respect to the usual
complex conjugation. However it turns out to be { \it real\/} with
respect to some combined conjugation, which is the product of the
complex and antipodal conjugations.

In this paper we restrict ourselves to a discussion of the self-dual
Yang-Mills theories in Euclidean space.
An outline of the paper is as follows. In section 2 we recall some
basic properties of harmonics and their relation to
quaternions. The necessity of complexification is discussed on the
simplest level in section 2.2. In section 3 the simplest notion of harmonic
analyticity is introduced.
Its r\^ole in self-dual gauge theories is
demonstrated in section 4, where also some other facts \cite{galp},
\cite{og}, \cite{sok} concerning the harmonic space treatment of
self-duality are collected, and simple examples of quaternionic
analytic transformations are given. Some other cosets of the 4D rotation
group (other than $S^2$) are discussed briefly in section 5, including
the one needed for the Fueter approach. Section 6 is devoted to a thorough
examination of the cosets of the conformal group $SO(5,1)$,
or, more precisely, of its universal cover $Spin(5,1)$, since we
have to deal with spinorial harmonics.
Again we
are compelled to consider its action on a coset, $CP^3$, of
its complexified form, $Spin(6,C) \sim SL(4,C)$. In this section
necessary techniques
are worked out, and we show how to calculate efficiently the form of
transformations on such cosets. Some mathematical definitions and
statements
are given in the Appendix, ``Compact cosets of non-compact groups".
Another Appendix presents a 5-parameter family of
quaternionic complex structures in 4D.

\section{ Harmonics}

\subsection{$SU(2)/U(1)$}

Before discussing these problems it will be worth recalling some basics
concerning harmonics \cite{galp}, \cite{gal}. As in any realization of
twistor program \cite{pen}, \cite{ward}, the harmonic space
approach \cite{gal} proceeds by considering an enlarged space,
which is the price paid to enable us to define appropriate
analyticities. In our case this space includes the
two-dimensional sphere $S^2$. We begin by considering it as a coset
of the rotation group $Spin(4)=SU(2)_L \times SU(2)_R$ modulo its subgroup
$U(1)_L \times SU(2)_R$. In other words, we present this sphere as a
coset  $SU(2)/U(1)$. Of course, one could choose polar ($\theta,\phi$)
or stereographic ($z,\bar z$) coordinates to describe
this sphere. However, it turns out to be much more convenient to
use just harmonics instead of any specific coordinates,
because harmonics are defined on the sphere {\it globally}.
We shall deal with a $2\times 2$ matrix \cite{gal}
\be
  U = (u^-_\alpha, u^+_\alpha)  =
\left(\begin{array}{cc} u^-_1 & u^+_1 \\u^-_2 & u^+_2 \end{array} \right)
\ee
Harmonics have
$SU(2)$ indices $\alpha$ and $U(1)$ charges $+,-$. They transform under
$SU(2)$ as spinors, thus for $M\in SU(2)$, $(M^+M=1)$, we have
\be
u^{\pm'}_\alpha = M_\alpha^\beta u^\pm_\beta,  \hskip 1cm
U' = MU.
\ee

In accordance with their coset nature, harmonics are
defined modulo $U(1)$ transformations, implemented by a matrix $P$,
\be
  U \longrightarrow U P,    \qquad\qquad
P=\left(\begin{array}{cc} e^{-i\lambda} & 0 \\
0 & e^{+i\lambda} \end{array} \right) \in U(1)      \label{s}
\ee
 or
\be
u^{+'}_\alpha \simeq e^{i\lambda}u^{+}_\alpha,\quad
u^{-'}_\alpha \simeq e^{-i\lambda}u^{-}_\alpha .
\label{S}
\ee

Owing to this freedom, one can take transformations of the
matrix $U$ in a form
\be
 U' = MUP \hskip 1 cm   (P^+P=1)   \label{mus}
\ee
Sometimes it is convenient to fix this matrix $P$ in a certain way in order
to pass to some
specific coordinates, etc. However, the global
description of a quotient manifold is then lost, and there arises the well
known Riemann-Hilbert problem.

Finally, in ${SU(2)\over U(1)}$ description of $S^2$, all matrices, $U$,
$M$ and $P$, are unitary, and harmonics of
opposite $U(1)$ charges are {\it complex conjugates},
\be
u^-_\alpha=\overline{u^{+\alpha}},    \label{compl}
\ee
where $SU(2)$ indices are raised in the usual way,
$u^{+\alpha} = \epsilon ^{\alpha \beta} u^+_\beta$.
Harmonics have to obey the constraint
\be
\det U = u^{+\alpha} u^-_\alpha = 1                     \label{5}
\ee
and the completeness relation
\be
u^{+\alpha} u^-_\beta - u^{-\alpha} u^+_\beta = \delta^\alpha _\beta
\label{con}        \ee
holds. This relation makes useful projection possible,
\be
f_\beta = (u^{+\alpha} u^-_\beta - u^{-\alpha} u^+_\beta) f_\alpha =
(u^{+\alpha} f_\alpha) u^-_\beta - (u^{-\alpha} f_\alpha) u^+_\beta,
\label{proj}  \ee
So all free undotted indices can be ascribed to harmonics only.
We shall often use this technique.

Note that since $2\times 2 $ matrices are unitary, both $M$ and
$U$ can be thought of as quaternions of unit norm.
Since harmonics are defined only up to $U(1)$, {\it the $U(1)$  phase
must not enter any formulae}. This means that the $U(1)$ charge
has to be conserved  and that ``functions'' on the
sphere must possess a {\em
definite} $U(1)$ charge, $q$. In other words, all terms in their
decomposition have to contain products of harmonics
$u^{+},u^{-}$ of the same charge $q$. For instance, for $q=+1$:
\be
f^{+}(u)=f^\alpha u^{+}_\alpha +
f^{\{\alpha \beta \gamma  \}}u_\alpha^{+}u_\beta ^{+}u_\gamma ^{-}
+\ldots \label{6}
\ee
Such quantities will acquire an overall $U(1)$ phase;
however, this is unimportant owing to the presupposed preservation
of the $U(1)$ charge. Of
course, complete symmetrization in indices $\alpha, \beta, \gamma \ldots$
is implied in each term of a harmonic
decomposition like (\ref{6}), which otherwise could be reduced to lower
order terms by using the constraint (\ref{5}).

   In fact, $u^+_\alpha , u^-_\alpha $ are the fundamental spherical
harmonics of
spin $1/2$, familiar to everyone from quantum mechanics%
\footnote{$ U = \left(\begin{array}{cc} \cos{\theta\over 2}
e^{-{i\phi\over 2}} & i\sin{\theta\over 2}e^{-{i\phi\over 2}}\\
i\sin{\theta\over 2}e^{i\phi\over 2} & \cos{\theta\over 2}
e^{{i\phi\over 2}}\end{array}\right)$ in Euler angles, the third
one is irrelevant, being the phase from eq. (12).},
while (16) is an example of the harmonic decomposition on $S^2$. This is
why we shall refer to $u^+_i,u^-_i$ in what follows simply as harmonics.

It is convenient to perform both differentiation and integration
on the two-sphere directly
in terms of harmonics. The action of the harmonic derivative $D^{++}$
on the harmonics themselves is defined according to a simple rule
\be
D^{++}u^+_\alpha=0,\hspace{1.5cm} D^{++}u^-_\alpha= u^+_\alpha.    \label{7}
\ee

\subsection{Complexification of $SU(2)$}

The following important note is in order. When considering conformal
invariance of the self-dual equations (as well as in some other cases)
it proves necessary to {\it complexify\/} the above treatment.
The reason is as follows. The parameters of conformal boosts have dimension
$length^{-1}$. So conformal transformations of harmonics will be
linear in the space coordinate. If $u^-, u^+$ were complex
conjugates, as in (\ref{compl}), and if the $u^+$ transformation were
analytic [see sections 3, 4 and (\ref{cb}), (\ref{conf})], then
transformations of $u^-$ would be unavoidably non-analytic. Considering
the action of $SU(2)$
on a coset of its complexification, one can
have both $u^+$ and $u^-$ transforming analytically (see sections 3, 4),
because in this case they cease to be complex conjugates of
one another.

Before, we presented the two-dimensional sphere $S^2$ as a coset
$SU(2)/U(1)$. Now we are going to consider the $SU(2)$ group action
in the $S^2$ coset of its complexification
$SL(2,C)$. The latter can be represented by $2\times 2$
unimodular matrices. It is non-compact and has a unimodular triangular
subgroup, which is its maximal parabolic subgroup \cite{vil},
\cite{bas}, \cite{hum} (see Appendix B for
mathematical definitions and techniques).
It is known that a two-sphere can also be considered as a coset of the
complexified group. Making use of the Iwasawa decomposition,
taking the parabolic subgroup
to be $P = U(1)\times A N$, $A$ and $N$ being subgroups of $SL(2,C)$, we have
$${SL(2,C)\over P}= {SU(2)\times A N \over U(1)\times A N} = {SU(2)\over
U(1)} = S^2,
$$   or
\be
S^2 = {SL(2,C)\over P} =
\frac{\left(\begin{array}{cc} a & b \cr
c & d \end{array} \right)}
{\left(\begin{array}{cc} \rho & 0  \cr
z^{--}  & \rho^{-1}  \end{array} \right)} , \hskip 0.5 cm
a d - b c = 1,           \label{par}
\ee
where $a, b, c, d , \rho$ and $z^{--}$ are complex
numbers. According to this presentation, harmonics are defined up to
the parabolic group transformations
\be
u^{+'}_\alpha = \rho^{-1}u^+_\alpha, \quad
u^{-'}_\alpha = \rho u^-_\alpha  + z^{--} u^+_\alpha, \label{++}
\ee
which are quite general for the $u^-$ harmonics. Consequently, we come to
the crucial conclusion that {\it any transformation of harmonics can be
reduced to}
\be
\delta u^+_\alpha = \lambda^{++} u^-_\alpha,\quad  \delta u^-_\alpha= 0,
\label{AA}
\ee
with some parameters $\lambda^{++}$ (choosing the appropriate
compensating parabolic group transformation).

We shall refer in what follows to this gauge fixing as the $u^-$ or
``normal"
form, because in all gauge theories, including gravity, there exists a
normal gauge in which all prepotentials depend only on $u^-$ and are
independent of $u^+$ harmonics. The normal form simplifies
reasonings and calculations considerably.

A general procedure for finding such a form follows from the
above rule (\ref{mus}).
For infinitesimal transformations $\delta M$ it reads:
\be
\delta U = \delta M U + U \Delta P  \label{inf}  \ee
One can always find such compensating
$\Delta P$ such that $\delta u^-_\alpha=0$. For instance, for rotations
one has
\be
\delta u^\pm_\alpha = \delta l_\alpha^\beta u^\pm_\beta.  \label{rot}
\ee
Passing to the $u^-$-form, we write
\be
(0, \delta u^+_\alpha) = (\delta l_\alpha^\beta u^-_\beta,
\delta l^\beta_\alpha u^+_\beta) + (u^-_\alpha \Delta \rho +
\Delta z^{--} u^+_\alpha, \ -\Delta \rho u^+_\alpha)
\ee
Then, projecting on harmonics [see (\ref{proj})], we obtain
for the parabolic transformation parameters
\be
\Delta \rho = -u^{+ \gamma} \delta l^\beta_\gamma u^-_\beta, \quad
 \Delta z^{--} = u^{- \gamma} \delta l^\beta_\gamma u^-_\beta,
\ee
while transformations of harmonics acquired the $u^-$ form
\be
\delta u^-_\alpha = 0, \hskip 0.7cm \delta u^+_\alpha =
(u^{+\gamma} \delta l_\gamma^\beta u^+_\beta) u^-_\alpha.     \label{lor}
\ee

With these new rules of the game {\it harmonics $u^+$ and $u^-$ are
no longer complex conjugates}. Nevertheless a new combined conjugation
can be defined \cite{gal}, which is a product of the complex conjugation
and the antipodal map (just a map of a point on the one end of diameter
to one on the other end):
\be
\hat u^\pm_\alpha = u^{\pm \alpha}, \qquad
\hat u^{\pm \alpha} = - u^\pm_\alpha.             \label{ccon}
\ee
The reality properties are discussed in terms of this newly
defined conjugation.

An important comment is that these reality properties  of harmonics are
preserved by the action of {\it $SU(2)$} on $S^2$,
but not by the complete $SL(2,C)$.

\subsection{Harmonics as square roots of quaternions}

We have mentioned above that harmonics are deeply related to quaternions.
In fact, in a general reference system, quaternions can be considered to be
bilinear combinations of harmonics.

To see this we unify $U(1)$ charges into one index $i$:
\be
u^\pm_\alpha = u_\alpha^i, \hskip15mm i =(+,-).
\ee
Then the defining constraint (\ref{5}) and completeness relation
(\ref{con}) acquire a symmetric form
\be
u^\alpha_i u^j_\alpha = \delta^i_j; \hskip15mm u^\alpha_i u^i_\beta =
\delta^\alpha_\beta.                \label{8}
\ee

Now the whole two-parameter family of quaternionic units that are
arbitrarily oriented in three-dimensional space is given by
\be
{e_a}_\alpha^\beta = -i u^i_\alpha {\sigma_a}^j_i u^\beta_j.  \label{9}
\ee
This can be easily checked with the help of (\ref{8}). The above
representation (\ref{2}) corresponds to a special gauge fixing
\be
u_1^i =\left( \begin{array}{c}1\\0\end{array}\right),\qquad
u_2^i =\left( \begin{array}{c}0\\1\end{array}\right).    \label{10}
\ee
This harmonic nature of quaternions explains why they are needed
in all problems where manifolds have quaternionic structures: in
the $N=2$ supersymmetric theories \cite{gal}, \cite{baggal} and
in the self-dual ones \cite{galp}, \cite{og}, \cite{sok}.

\section{Harmonic quaternionic analyticity}

We shall begin our discussion of quaternionic analyticity by recalling
some arguments of \cite{galp}.
First of all, speaking of the coordinate $x^{\alpha \dot\alpha}$ as a
quaternion $z$, one has in mind the 4D rotation group in
the form $Spin(4) =
SU(2)_L\times SU(2)_R$. It is natural to realize it on some of its coset
spaces. The simplest possibility is to choose two-sphere
\be
SU(2)_R/U(1) = \{u^\pm_\alpha\}                \label{11}
\ee
It is convenient to pass to space coordinates
\be                                                            \label{12}
x^{\pm\dot\alpha} = x^{\alpha \dot\alpha} u^\pm_\alpha, \hskip15mm
x^{\alpha \dot\alpha} = -x^{+\dot\alpha}u^{-\alpha}+
x^{-\dot\alpha}u^{+\alpha}
\ee
As one can recognize, this is the Penrose twistor transformation,
written in the language of harmonic space. The usage of $x^+$ and
$x^-$ coordinates permits the introduction of a new
kind of analytic function, which is dependent on $x^+$ and
harmonics, but is
independent of $x^-$. The corresponding Cauchy-Riemann conditions will be
of first order in derivatives:
\be
{\partial\over\partial{x^{-\dot\alpha}}} f^A(x,u)=0,       \label{An}
\ee
where $A$ symbolizes the spinor indices and the $U(1)$ charges. As a
consequence of this condition we have
\be
\Box f^A = {\partial\over\partial{x^m}}\times
{\partial\over\partial{x^m}} f^A =0   \label{f}
\ee
because one can check that $\Box=
{\partial\over\partial{x^{+\dot\alpha}}}\times
{\partial\over\partial{x^-_{\dot\alpha}}} $ is the usual d'Alembertian
in four dimensions. So
the d'Alembertian of quaternionic analytic function vanishes for
reasons completely analogous to those acting in the case of the customary
complex analyticity.

It is evident that the property of analyticity is preserved
by the general quaternionic analytic transformations mixing coordinates
$x^{+\dot\alpha}, u^+_\alpha, u^-_\alpha$ :
\be
\delta x^{+{\dot\alpha}} = f^{+\dot\alpha}(x^+, u^\pm),   \label{13}
\ee
$$
\delta u^+_\alpha = w^{++}(x^+, u^\pm) u^-_\alpha
$$
$$
\delta u^-_\alpha = 0
$$
with arbitrary quaternionic analytic functions $f^{+\dot\alpha}$ and
$w^{++}$ as parameters. When writing down these transformations, we
have effectively taken into account that harmonics are defined modulo
transformations (\ref{++}). Note also that no
assumptions were made concerning a form of transformations of coordinates
$x^{-\dot\alpha}$,
\be
\delta x^{-\dot\alpha }=\phi^{-\dot\alpha}(x^+, x^-, u^\pm),\\ \label{13a}
\ee
where local parameters $\phi$ are non-analytic and can depend on
$x^-$ in any manner. To be more concrete, we shall give some examples.

1. The Poincar\'e group (left and right rotations,
$\delta l^\alpha_\beta$
and $ \delta r^{\dot\alpha}_{\dot\beta}$, respectively,
$\delta l^\beta_\beta= \delta r^{\dot\alpha}_{\dot\beta} = 0 $,
and translations
$\delta b^{\alpha \dot\alpha}$) is represented by analytic quaternionic
transformations:
$$
\delta x^{+{\dot\alpha}} =
- \delta r^{\dot\alpha}_{\dot\beta} x^{+{\dot\beta}}
 + \delta b^{\alpha \dot\alpha} u^+_\alpha, \quad
\delta u^\pm_\alpha = +\delta l^\beta_\alpha u^\pm_\beta,
$$
$$
(\delta x^{\alpha \dot\alpha } =
-\delta l^\alpha_\beta x^{\beta \dot\alpha} -
\delta r^{\dot\alpha}_{\dot\beta} x^{\alpha \dot\beta} +
\delta b^{\alpha \dot\alpha}, \quad
\delta x^{-{\dot\alpha}} =
- \delta r^{\dot\alpha}_{\dot\beta} x^{-{\dot\beta}} +
\delta b^{\alpha\dot\alpha} u^-_\alpha).
$$
Note that rotations of harmonics can be also represented as in (\ref{rot}).

2. The same for dilatations
$$
\delta x^{+{\dot\alpha}} = \delta d x^{+{\dot\alpha}},\quad
\delta u^\pm_\alpha =0
$$
and $\delta x^{\alpha \dot\alpha} = \delta d  x^{\alpha \dot\alpha},
\quad \delta x^{-{\dot\alpha}} = \delta d  x^{-{\dot\alpha}}.$
[It must be remembered
that harmonics are defined modulo transformations (\ref{++}).]
The above transformations exhaust all
quaternionic analytic transformations that are
linear in $x^{+{\dot\alpha}}$ and do not lead to an explicit
appearance of harmonics in $\delta x^{\alpha \dot\alpha}$.

3. The affine transformations
$$
\delta x^{\alpha \dot\alpha} = a^{\alpha \dot\alpha}_{\beta \dot\beta}
x^{\beta \dot\beta}, \quad  a^{\alpha \dot\alpha}_{\alpha \dot\beta}=
a^{\alpha \dot\alpha}_{\beta \dot\alpha} = 0
$$
are definitely non-analytic for any choice of reparametrizations of
harmonics. For instance, if $\delta u^\pm_\alpha=0$, then
$$
\delta x^{+ \dot\alpha} = u^+_\alpha
a^{\alpha \dot\alpha}_{\beta \dot\beta} (u^{+\beta} x^{-\dot\beta}-
u^{-\beta} x^{+\dot\beta})
$$
contains both $x^{+ \dot\alpha}$ and $x^{- \dot\alpha}$.

4. Passing to transformations bilinear in $x^{+ \dot\alpha}$ we observe,
first of all, that {\it the conformal boosts belong to the quaternionic
analytic transformations}. Explicitly, conformal boosts are,
\be
\delta x^{\alpha \dot\alpha} =
x^{\alpha \dot\beta} \delta k_{\dot\beta \beta} x^{\beta \dot\alpha}
\label{cb}
\ee
($\delta k_{\dot\beta \beta}$ are parameters), so that
\be
\delta x^{+ {\dot\alpha}} =- x^{+ {\dot\beta}} \delta k_{\dot\beta \beta}
u^{-\beta} x^{+{\dot\alpha}},           \label{co}
\ee
\be
\delta u^+_\alpha = - x^{+\dot\beta} \delta k_{\dot\beta \beta}
u^{+\beta} u^-_\alpha, \quad \delta u^-_\alpha = 0,     \label{conf}
\ee
$$
(\delta x^{-{\dot\alpha}} =- x^{+ {\dot\beta}} \delta k_{\dot\beta \beta}
u^{-\beta} x^{-{\dot\alpha}} ).
$$
The conformal boost parameter has dimension
$length^{-1}$. So, conformal transformations for harmonics contain
space coordinates $x$ linearly. It is worth stressing once again
that it becomes possible to avoid an appearance of $x^{- \dot\alpha}$
(non-analyticity) in the transformations of $u^-_\alpha$ harmonics only
because any transformation of $u^-$ can be compensated
by an appropriate parabolic group one (passing to the normal form).

Note that under combined conjugation (\ref{ccon})
\be
\hat x^{+ \dot\alpha} = - x^+_{\dot\alpha}    \label{cconx}
\ee
and it is consistent with {\it real} Poincar\'e and conformal
transformations.

Note also that reality properties of the above general transformations
(\ref{13}) have to be consistent with the combined conjugation
(\ref{ccon}), (\ref{cconx}) as well.

We shall restrict ourselves to these  examples.

It is remarkable that {\it just this kind of analyticity is inherent} in
the important theories discussed in the next section.

\section{Self-dual gauge theories in four dimensions}

To recover the quaternionic  analytic nature of the theories quoted in the
title, we shall recall here the Ward procedure \cite{ward}, \cite{pen}
for dealing with
the self-dual gauge equations in $R^4$. The usage of
the harmonic space language \cite{galp}, \cite{og},
\cite{sok} makes the situation completely understandable, showing a
transparent correspondence between a quaternionic analytic (in the sense
of the preceding section) double $U(1)$ charged function $V^{++}(x^+,u)$
and solutions of the self-dual equations.
Note that consideration of the self-dual Einstein theory goes
along similar lines.

The commutator of covariant derivatives
$D_{\alpha \dot\alpha}= \partial_{\alpha \dot\alpha}+
i A(x)_{\alpha \dot\alpha}$ (connection $A$ takes values in the Lie
algebra of the gauge group for the Yang-Mills theory and in the tangent
Lorentz group for the Einstein theory),
\be                                                       \label{16}
[D_{\alpha \dot\alpha},D_{\beta \dot\beta}] = \epsilon_{\alpha \beta}
F_{\dot\alpha \dot\beta}(x) +\epsilon_{\dot\alpha \dot\beta} F_{\alpha
\beta}(x),
\ee
defines the Yang-Mills field-strengths, $F_{\alpha \beta}, F_{\dot\alpha
\dot\beta}$.
In the spinor formalism, the self-duality equation is simply
$ F_{\dot\alpha \dot\beta}(x)=0$.  Taking into account the definition
(\ref{16}), it is evident that this equation is {\em equivalent to}
\be
[D_{\alpha \dot\alpha},D_{\beta \dot\beta}] =
\epsilon_{\dot\alpha \dot\beta} F_{\alpha \beta}(x) . \label{17}
\ee
Having harmonics at our disposal, we can disentangle this equation
in the following way.
Multiplying Eq. (\ref{17}) by $u^{+\dot\alpha} u^{+\dot\beta}$, and
defining
\be
D^+_\alpha = u^{+\dot\alpha} D_{\alpha \dot\alpha},\label{m}
\ee
we obtain
\be
[D^+_\alpha,D^+_\beta] = 0.                          \label{18}
\ee
It is more convenient to replace (\ref{m}) with the equivalent (taking
into account $U(1)$ charge conservation) commutator relation (cf. (18))
\be
[D^{++},D^+_\alpha]=0.                \label{19}
\ee

{\em The pair of equations (\ref{18}) and (\ref{19}) is equivalent to the
self-duality condition (\ref{17})}. However this pair is considerably simpler.
The first of them states that the covariant derivatives $D^+$ commute.
So its solution is ``pure gauge'',
\be
D^+_\alpha= h \partial^+_\alpha h^{-1}= \partial^+_\alpha +
h (\partial^+_\alpha h^{-1}),     \label{20}
\ee
where derivative $\partial^+_\alpha$ does not contain any connection, and
a ``bridge" $h=h(x,u)$ takes values in the gauge group.
By choosing coordinates (\ref{12}) one has
\be
D^+_\alpha=u^{+\dot\alpha}\partial_{\alpha \dot\alpha}={\partial\over{
\partial{x^{-\alpha}}}}.        \label{21}
\ee
In the gauge where $D^+_\alpha$ becomes short, the harmonic derivative $D^{++}$
grows long (because the bridge $h$ generally depends on harmonics):
\be
D^{++}\longrightarrow {\cal D}^{++}= h^{-1}D^{++} h=
D^{++} + h^{-1} (D^{++} h) = D^{++} + i V^{++},                \label{22}
\ee
acquiring a harmonic connection that is globally defined on $S^2$
(gauge algebra valued indeed)
\be
V^{++}=-ih^{-1}\left(D^{++} h\right) .                \label{23}
\ee
Now the second equation of the pair becomes the {\em Cauchy-Riemann
condition for the harmonic analyticity }
\be
{\partial\over {\partial x^{-\alpha}}} V^{++} = 0,     \label{24}
\ee
stating that for the self-dual case the harmonic connection is
analytic, i.e. it is independent of
$x^{-\alpha}$, $V^{++}=V^{++}(x^{+\alpha},u^\pm)$. Vice versa,
if $V^{++}$ is analytic, it encodes a solution of the self-dual
equations.

Moreover, one can get rid of positively charged harmonics $u^+$,
since gauge potentials $V^{++}$ are defined only
up to gauge transformations
$$ \delta V^{++}(x^+,u) = (D^{++} + i V^{++}(x^+,u)) \lambda(x^+,u).
$$
There exists a {\it normal gauge\/} \cite{galp} where
$V^{++}$ contains in its harmonic decomposition only negatively charged
harmonics $u^-$:
$$V^{++} = V^{++}(x^+,u^-).$$

So, an analytic $V^{++}$ encodes a solution of the self-dual equation.
This is a transparent manifestation of the ``twistor correspondence".
However we are more familiar with the usual space $R^4$ than with
the harmonic one. To pass to the former, one has to determine the bridge
$h$ from equation (\ref{23}) for a given analytic $V^{++}$, and
then substitute the bridge $h$ into the expression for the
usual vector connection $A_a(x) =-i h{\partial\over\partial{x^a}} h^{-1}$.
Of importance is that (\ref{23}) has a solution for almost
any $V^{++}$ \cite{og}.

Thus, there is no problem in solving the self-duality equations in the
harmonic space, while a solution of (\ref{23}) on the two-sphere
is needed to
pass to the ordinary space. Instantons and monopoles are special
solutions of the self-duality equation having finite action and finite
energy, respectively. They have been completely described in the harmonic
space language, including ADHM construction, etc. \cite{{og},{sok}}.

Of course, any change of the harmonic analytic connection
\be
V'(x^+,u) = V^{++}(x^+,u) + g^{++}(x^+,u)
\ee
results in passing from one solution of the self-dual equation to
another. So the most general B\"acklund transformation is encoded
again by the double $U(1)$ charged analytic object $g^{++}(x^+,u)$. An
important geometric class of them consists of the general analytic
diffeomorphisms (29)
accompanied with a ``similarity" transformation, defined by a
 general analytic weight $c(x^+,u)$ that takes values in the gauge
algebra,
\be
V'(x^{+'},u') = e^{c(x^+,u)} V^{++}(x^+,u) e^{-c(x^+,u)}.   \label{ba}
\ee
This class includes a great many
B\"acklund transformations. Speaking of
diffeomorphisms, we can restrict ourselves to those that are realized
in the normal form.

So, a kind of quaternionic analyticity, the harmonic one, is
inherent in the self-dual 4D gauge theories. Of course, they are also
conformally invariant. However, now conformal transformations form a
finite-dimensional subgroup $Spin(5,1)$ of analytic transformations, just
those given by (\ref{co}), (\ref{conf}). Therefore these theories
can be naturally  considered as 4D-extensions of the 2D-conformal
field theories in both their conformal and analytic aspects.

\section{Examples of other quaternionic analyticities}

Only one coset space was considered above, just
the two-sphere $S^2$.
There are other possibilities, however, some of which we shall
now discuss briefly.

\subsection{A product of two $S^2$}

  The first example is
\be
\frac {SU(2)_L} {U(1)_L} \times \frac {SU(2)_R} {U(1)_R}
 \ee
with harmonization of both the left and right $SU(2)$ groups and with two
distinct $U(1)$ charges. In this case there are both left
($v^{\oplus,\ominus}_\alpha$) and right ($u^{+,-}_{\dot\alpha}$) harmonics
having left ($\oplus,\ominus$) or right ($+,-$) $U(1)$ charges,
respectively. The definition of the corresponding quaternionic analyticity
is rather obvious. One has to split $x^{\alpha \dot\alpha}$ into
four pieces
\be               \label{D}
x^{+\oplus}=x^{\alpha \dot\alpha}v^\oplus_\alpha u^+_{\dot\alpha}, \hskip10mm
x^{+\ominus}=x^{\alpha \dot\alpha}v^\ominus_\alpha  u^+_{\dot\alpha},
\hskip10mm
x^{-\oplus}=x^{\alpha \dot\alpha}v^\oplus_\alpha u^-_{\dot\alpha}, \hskip10mm
x^{-\ominus}=x^{\alpha \dot\alpha}v^\ominus_\alpha u^-_{\dot\alpha}.
\ee
It is easy to arrange quaternion conjugations that transform these variables
among themselves. To combine them into the ordinary 4-coordinate is also
straightforward:
\be
x^{\alpha \dot\alpha}= v^{\oplus \alpha} u^{+ \dot\alpha} x^{-\ominus}
-v^{\oplus \alpha} u^{- \dot\alpha} x^{+\ominus}
-v^{\ominus \alpha} u^{+ \dot\alpha} x^{-\oplus}
+v^{\ominus \alpha} u^{- \dot\alpha} x^{+\oplus}.
\ee
One can define analytic functions to depend on only
one of the four coordinates (\ref{D}), say on $x^{-\oplus}$ (and in some
way on harmonics). Then there will be three Cauchy-Riemann conditions of
the first order in derivatives:
\be
\partial^{+\oplus}f=\partial^{-\ominus}f=\partial^{-\oplus}f=0.
\ee
Again, the consequence is that the d'Alembertian of the analytic
function vanishes: $\Box f \equiv \partial^{+\oplus}\partial^{-\ominus}f-
\partial^{-\oplus}\partial^{+\ominus}f= 0 $.
At present we do not know a field-theoretical model connected with such
quaternionic analyticity.

\subsection{A diagonal $U(1)$ case}

Further, one can {\em identify} two $U(1)$ groups and consider the coset
space
\be
\frac {SU(2)\times SU(2)} {U(1)}
\ee
that is connected with the same harmonics
$v^\pm_\alpha, u^\pm_{\dot\alpha}$
as in the above example, both having, however, charges of the same
diagonal $U(1)$ subgroup.
This circumstance, as we shall see, will help. As in the previous case,
the four-coordinate is split up into four separate variables,
\be
x^{++}=x^{\alpha \dot\alpha}v^+_\alpha u^+_{\dot\alpha}, \hskip5mm
x^1 =x^{\alpha \dot\alpha}v^-_\alpha  u^+_{\dot\alpha}, \hskip5mm
x^2=x^{\alpha \dot\alpha}v^+_\alpha u^-_{\dot\alpha}, \hskip5mm
x^{--}=x^{\alpha \dot\alpha}v^-_\alpha u^-_{\dot\alpha}, \label{59}
\ee
while
\be
x^{\alpha \dot\alpha}= v^{+ \alpha} u^{+ \dot\alpha} x^{--}
-v^{+ \alpha} u^{- \dot\alpha} x^1 -v^{- \alpha} u^{+ \dot\alpha} x^2
+v^{- \alpha} u^{- \dot\alpha} x^{++}.
\ee
The Cauchy-Riemann conditions are again of the first order in
derivatives; for a function that has to depend only on,
say, $x^1$ they are
\be
\partial^{++} f= \partial^2 f = \partial^{--} f = 0  \label{B}
\ee
and  any analytic function satisfying (\ref{B}) will obey
 \be
\Box f\equiv \partial^{++}\partial^{--}f-\partial^1\partial^2f= 0,
\ee
where $\partial^1$ and $\partial^2$ differentiate with respect to
$x^1$ and $x^2$, respectively.

\subsection{Fueter quaternionic analyticity revisited}

As was stated in the introduction, a Fueter analytic function does
not satisfy the Cauchy-Riemann condition or the equation
$\Box f = 0$. Instead the Cauchy-Riemann-like condition holds
for $\Box f$, leading to an equation of fourth order,
$\Box^2 f = 0$. To demonstrate these statements
it is worth emphasizing that {\it Fueter analyticity is connected
just to the harmonic approach of section 5.2}. Indeed, in \cite{gj} it
was shown that to make the Fueter decomposition (\ref{F}) formally
covariant one has to introduce another quaternion $p^{-1}$, with the
transformation properties inverse to those of  $z$:
\be
z'= m z\bar n, \quad  p^{-1'}=np^{-1}\bar m, \quad  m\bar m = n\bar n =1,
\ee
where $m\in SU(2)_L$ and $n\in SU(2)_R$ are unit quaternions
representing these groups, respectively (cf. footnote 5)

Now a new variable, termed a left quator in \cite{gj}, \cite{gurs}
\be
y=zp^{-1}
\ee
will have a ``more convenient", purely left transformation law
\be
y'= m y \bar m.
\ee
Correspondingly, the modified Fueter definition \cite{gurs}
\be
f(y)=\sum {a_n y^n} \label{66}
\ee
will be consistent with four-dimensional rotations, with $y$ belonging
to the $(1,0)\oplus (0,0)$ representation of $SO(4)$.
It is easy to see
that this newly
introduced auxiliary quaternion $p$ {\it may be taken to be a vector
harmonic}, composed of the spinor harmonics introduced in
section 5.2 according to
\be
p^{-1}_{\alpha \dot\alpha}= (v^+_\alpha u^-_ {\dot\alpha}
+v^-_\alpha u^+_{\dot\alpha}).                            \label{E}
\ee
[There is some freedom in defining the coefficients
on the right-hand side
of (\ref{E})]. Note that $p^{-1}$ becomes the unit matrix in the special
reference system (\ref{10}) for both types of harmonics.  Thus the Fueter
analytic function is a power series  of a left quator (in the
terminology of \cite{gurs})
\be
y^\alpha_\beta = x^{\alpha\dot\alpha}(v^+_\beta u^-_{\dot\alpha} +
v^-_\beta u^+_{\dot\alpha}) =
x^{++}(L^{--})^\alpha_\beta + x^{--}(L^{++})^\alpha_\beta +
x^1(P^1)^\alpha_\beta + x^2(P^2)^\alpha_\beta,  \label{quator}
\ee
where definitions (\ref{59}) were used, and we have introduced operators
\begin{eqnarray}
(L^{++})^\alpha_\beta &= v^{+\alpha}v^+_\beta,\qquad\qquad
(L^{--})^\alpha_\beta &= -v^{-\alpha}v^-_\beta, \nonumber\\
\mbox{}&\mbox{}&\mbox{}\\
(P^1)^\alpha_\beta &= v^{+\alpha}v^-_\beta,  \qquad\qquad
(P^2)^\alpha_\beta &= -v^{-\alpha}v^+_\beta.    \nonumber
\end{eqnarray}
They satisfy the following algebra (with indices suppressed for brevity):
the operators $P$ are projectors,
\be
P^1 P^1 = P^1,\qquad P^2 P^2 = P^2,\qquad P^1 + P^2 = 1,\qquad
P^1 P^2 = P^2 P^1 = 0, \label{70}
\ee
while for the $L$'s we have
\be
L^{--} L^{++} = P^1,\qquad L^{++} L^{--} = P^2,\qquad L^{--} L^{--} =
L^{++} L^{++} = 0 \label{71}
\ee
and the remaining products are
\begin{eqnarray}
\mbox{}&L^{++} P^1 = P^2 L^{++} = L^{++},\qquad
L^{--} P^2 = P^1 L^{--} = L^{--},&\mbox{}   \nonumber\\
\mbox{}&\mbox{}&\mbox{}\\
\mbox{}&L^{++} P^2 = P^1 L^{++} = L^{--} P^1 = P^2 L^{--} = 0.
&\mbox{}\nonumber
\end{eqnarray}

We now wish to show that the general Fueter-analytic function of equation
(\ref{66}) is biharmonic, and satisfies some equation of third order in
derivatives. To this end, consider an integral representation for $f$,
\be
f(y) = \frac{1}{2\pi i} \oint dz \frac{f(z)}{z-y}  \label{73}
\ee
which follows from the fact that it has a Weierstrass-like decomposition
(\ref{66}). Thus we may restrict out attention to the function
$(z-y)^{-1}$, or, simpler yet, $y^{-1}$. Using the above algebra we
have
\be
y^{-1} = (x^{++} x^{--} - x^1 x^2)^{-1} z \label{74}
\ee
where
\be
z = x^{++}(L^{--}) + x^{--}(L^{++}) - x^1(P^2) - x^2(P^1)
\label{75}
\ee

It is now helpful to consider the differential operators,
\begin{eqnarray}
V &=& L^{--} \partial^{++} + L^{++} \partial^{--}
        -P^1\partial^1 - P^2\partial^2, \nonumber\\
\mbox{}&\mbox{}&\mbox{} \label{76}\\
T &=& L^{--} \partial^{++} + L^{++} \partial^{--}
        + P^1\partial^2 + P^2\partial^1,        \nonumber
\end{eqnarray}
which satisfy
\begin{eqnarray}
\mbox{}&VT = TV = \partial^{++}\partial^{--} - \partial^1 \partial^2 =
\Box,&\mbox{}\label{77}\\
\mbox{}&\mbox{}&\mbox{}\vphantom{(}\nonumber\\
\mbox{}&Tz = 0.&\mbox{} \label{78}
\end{eqnarray}
Taking into account the fact that
\be
\Box (x^{++} x^{--} - x^1 x^2)^{-1} = \Box (1/x^2) = 0  \label{79}
\ee
we see from (\ref{77})--(\ref{79}) that
\be
V^2 T y^{-1} = 0.       \label{80}
\ee
Thus we have proven that any Fueter-analytic function $f(y)$
satisfies the third order, ``Cauchy-Riemann," condition
\be
(L^{--} \partial^{++} + L^{++} \partial^{--} - P^1\partial^1 -
        P^2\partial^2) \Box f = 0       \label{81}
\ee
and hence the biharmonic equation
\be
\Box^2 f = 0.   \label{82}
\ee

It is worth recalling that the conformal group
of Euclidean 4-dimensional space
can be represented by Fueter-type transformations \cite{gj}. They are
realized on $z$ non-linearly as quaternionic-M\"obius transformations
\cite{oldgu} constructed in the usual way from the quaternionic entries
of the two-by-two matrix belonging to $SL(2,H)$.
\be
z' = (az+b)(cz+d)^{-1},     \label{mo}
\ee
where $a$, $b$, $c$ and $d$ are constant quaternions, satisfying
\be
\det (a - bd^{-1}c)\times \det d= | a d - b d^{-1} c d |^2= 1 \label{det}
\ee
(the unimodularity condition for a $2\times2$ matrix with quaternionic
entries)\footnote{A $2\times 2$ matrix with quaternionic
(or again $2\times 2$ matrix) entries can be decomposed into a product
$$
\left (\begin{array}{cc} a & b \cr c & d \end{array}\right) =
\left (\begin{array}{cc} I & bd^{-1} \cr 0 & I \end{array}\right)
\left (\begin{array}{cc} a - bd^{-1}c & 0 \cr 0 & d \end{array}\right)
\left (\begin{array}{cc} I & 0 \cr d^{-1}c & I \end{array}\right)
$$
of matrices having evident determinants \cite{gj}.}.
Indeed, if $c \neq 0, d \neq 0$, then $z'$ can be written
as $$ z' = a c^{-1} + (b d^{-1} - a c^{-1}) (1 + c z d^{-1})^{-1} $$
and it is a sum of a constant quaternion and a Fueter analytic function
of a composite argument (involving transformation parameters besides
the coordinate itself)
$y = c z d^{-1}$ multiplied from the left by another constant
quaternion. If $c = 0$, then $d \neq {0}$ and $z'$ is a linear function
of $y = a z d^{-1}$. Finally, for $d =0, c \neq {0}$, $z'$ is a linear
function of $t = b z^{-1} c^{-1}$.
Four-dimensional rotations correspond to (\ref{mo}) with $b=c=0$, $a=m$
and $d=n$, $m\bar m = n\bar n = 1$, cf. (8); dilatation is
generated when $ a$ is a real parameter, $d = 1, b=c=0$; translations
have parameter $b$ while $a = d = 1, c = 0 $; for conformal boosts $c$
is a parameter and $a = d = 1, b = 0$

In \cite{gj} infinite-dimensional quasi-conformal groups are
considered that generalize (\ref{mo}), being subgroups of the
four-dimensional group of diffeomorphisms.

\section{Unifying space and two-sphere. Complexifying the conformal
group}

Above we harmonized the rotation group $SO(4)$, and harmonics came out
without a visible connection to the space coordinates, $x^m$, that are
coordinates of the coset of the Poincar\'e group modulo its rotation
subgroup $SO(4)$. It would be desirable to have space and harmonic
(twistor) coordinates treated on an equal footing. Conformal symmetry
helps us achieve this goal.

The conformal group for the Euclidean 4-dimensional space is well-known
to be $SO(5,1)$, however, since we are dealing with
harmonics in spinorial representations of the Lorentz group, we are
really dealing with its universal cover, $Spin(5,1)$.
One could start by considering its cosets to find out
whether there is a suitable six-dimensional one. In the previous
section the quaternion-M\"obius form of $SO(5,1)$ was mentioned. In
spinor form it is represented by a matrix
\be
M = \left ( \begin{array}{cc} l^\alpha_\beta  & b^{\dot\beta}_\alpha \cr
c^\beta_{\dot\alpha} & r^{\dot\beta}_{\dot\alpha}   \end{array} \right )
\label{O5}       \ee
with unit determinant
\be
\det (l^\beta_\alpha - b^{\dot\alpha}_\alpha
(r^{-1})^{\dot\beta}_{\dot\alpha} c^\alpha_{\dot\beta}) \times
\det r^{\dot\beta}_{\dot\alpha} = 1
\ee
(see footnote 7). It has the same  entries as in section 5:
$l^\alpha_\beta$ and $r^{\dot\alpha}_{\dot\beta}$ present left and right
rotations, respectively, and dilatations, while $b^{\dot\beta}_\alpha$
and $c^\beta_{\dot\alpha}$ are translations and conformal boosts,
respectively. Now using the Iwasawa decomposition
\be
Spin(5,1) = Spin(5)\times A N       \label{51}
\ee
(see Appendix B) we really can find a six-dimensional
coset. To this end, one has to choose $Spin(3)\times SO(2)\times A N$ as
a parabolic subgroup $P$ (the same $A,N$  as in (\ref{51})).
Then the Grassmanian
\be
{Spin(5,1)\over P} = {SO(5)\over SO(3)\times SO(2)}   \label{SO5}
\ee
will be the only 6-dimensional coset.
However the left rotation group $SU(2)_L$ comes
out in the original {\it non-complexified form}. By the same argument as
in section 2, this will lead to
non-analytic conformal transformations.

This again suggests {\it complexification}, now of {\it the conformal
group}. We are therefore led to consider the action of $Spin(5,1)$ on
cosets of $Spin(6,C) \sim SL(4,C)$. Indeed, this works. Starting with
the Iwasawa
decomposition (see  Appendix B) of the latter,
\be
SL(4,C) = SU(4) \times A N,    \label{o6}
\ee
 and choosing the parabolic subgroup to be
\be
P = SU(3)\times U(1) \times A N
\ee
(with the same $A N$ as in (\ref{o6})) we come to the coset
\be
{SL(4,C) \over P} = {SU(4) \over SU(3) \times U(1)} = CP^3 \label{cp}
\ee
the $A N$ in the numerator and denominator being ``cancelled". Note that
the appearance of the 3-dimensional complex projective manifold
agrees with the twistorial literature \cite{pen}, \cite{penr}, \cite{ward},
etc. Coordinates of this manifold are two space coordinates
$x^{+\dot\alpha}$ and harmonics that can be represented by one
complex coordinate, as we shall now see.

 The reader can consult Appendix B for some definitions and techniques.
Using them, we shall give here a direct derivation in brief of the
form of the $Spin(5,1)$ transformations realized on this $CP^3$ coset.
We shall proceed in the same manner as we did in section 2, where
we dealt with the appropriate coset of the complexified $SU(2)_L$ group.

Generally speaking, it is better to work with the full set of
harmonics forming a $Spin(5,1)$ matrix
$$
U = \left (\begin{array}{cc} u_\alpha^s & u_\alpha ^{\dot s}   \cr
 u^s_{\dot\alpha} & u^{\dot s}_{\dot\alpha} \end{array} \right ).
$$
and identify them under the action of the subgroup $P$. This would be a
{\it global} definition of $G/P$, and within this framework the
Riemann-Hilbert problem would be completely avoided, etc.

However to show how to work just with 6 ordinary coordinates of the
6-dimensional manifold, we shall use here the subgroup $P$ to eliminate
locally redundant degrees of freedom in $U$. These local
coordinates of our coset can be written as the entries of a triangular
matrix:
\be
U = \left (\begin{array}{cc} u_\alpha^s & -u^-_\alpha x^{+\dot s}   \cr
 0 & \delta^{\dot s}_{\dot\alpha} \end{array} \right ); \label{M}  \ee
The conformal group acts on $U$ by multiplication from the left by a matrix
$M\in Spin(5,1)$, equation (\ref{O5}). To preserve the form (\ref{M})
we fix a gauge by using appropriate compensating parabolic
group transformations
$P$ [cf. (11) and (\ref{inf})]. For infinitesimal transformations we have
\be
\delta U = M \times U \times P - U \approx \delta M \times U +
U \times \Delta P         \label{dM}
\ee
As in section 2, harmonics are defined only up to a transformation
(\ref{++}) belonging to the parabolic group. So
we can take as a starting point that a gauge (\ref{AA})
is fixed, i.e. we shall work with transformations in the normal form,
$$
\delta u^-_\alpha = 0, \quad  \delta u^+_\alpha =
\lambda^{++} u^-_\alpha.     $$

We shall now calculate explicitly the transformations of $x^{+\dot\alpha}$
and $u^+_\alpha$, as well as the compensating transformations belonging to
the parabolic group, using
equation (\ref{dM}) together with the requirement (\ref{AA}). The
ingredients are

\be
A.\quad \delta U = \left (\begin{array}{cc}(0, \lambda^{++} u^-_\alpha )&
-u^-_\alpha \delta x^{+\dot p}  \cr 0 & 0 \end{array} \right ), \label{U}
\ee
\be
B. \quad  \delta M \times U = \left (\begin{array}{cc}
\delta {\tilde l^\beta_\alpha} u_\beta^s + \delta d \delta^s_\alpha  &
-\delta {\tilde l^\beta_\alpha} u_\beta^- x^{+\dot s} -
\delta d  u^-_\alpha x^{+ \dot s} + \delta b_\alpha^{\dot s}\cr
\delta c^\beta_{\dot\alpha} u^s_\beta &
-\delta c^\beta_{\dot\alpha} u^-_\beta x^{+\dot s} +
\delta {\tilde r^{\dot s}_{\dot\alpha}} -
\delta d \delta^{\dot s}_{\dot\alpha} \end{array} \right ),  \label{MU}
\ee
where we singled out dilatations $\delta d$: now
$\delta \tilde l^s_s=\delta \tilde r^{\dot s}_{\dot s} = 0$.

The induced parabolic group transformations form a matrix
\be
 \Delta P = \left (\begin{array}{cc} \left (\begin{array}{cc}
\Delta a + \Delta d & 0 \cr  \Delta z^{--}  &  -\Delta a + \Delta d \cr
  \end{array}  \right )_p^s     &
\left (\begin{array}{c} 0 \cr  \Delta a^{- \dot s}
\end{array}\right)_p \cr
\Delta c^s_{\dot p}     &  -\Delta d  \delta^{\dot s}_{\dot p} +
\Delta \tilde r^{\dot s}_{\dot p} \end{array}\right)
\ee
For the last ingredient, the matrix $U \times \Delta P$, we shall write
its entries separately:
The upper left corner:
\be
(u^-_\alpha (\Delta a + \Delta d - x^{+ \dot p} \Delta c^-_{\dot p}) +
u^+_\alpha \Delta z^{--}, \ u^+_\alpha (-\Delta a + \Delta d) -
u^-_\alpha x^{+\dot p} \Delta c^+_{\dot p}).  \label {C1}
\ee
The lower left corner:
\be
\Delta c_{\dot\alpha}^s.                                  \label {C2}
\ee
The upper right corner:
\be
u^+_\alpha \Delta a^{-\dot s} + u^-_\alpha \Delta d x^{+\dot s}
- u^-_\alpha x^{+\dot p} \Delta \tilde r^{\dot s}_{\dot p}.  \label {C3}
\ee
The lower right corner:
\be
- \Delta d \delta^{\dot s}_{\dot \alpha}
+ \Delta \tilde r^{\dot s}_{\dot\alpha}.   \label {C4}
\ee
Now we have to substitute ingredients (\ref{M}), (\ref{MU}), (\ref{C1}),
(\ref{C2}), (\ref{C3}) and
(\ref{C4}) into equation (\ref{dM}). Then Projecting all entries on $
u^\pm_\alpha$, we get from the resulting equations explicit expressions
for the infinitesimal transformations of coset coordinates:
$$
\delta x^{+\dot\alpha} = ( 2 \delta d +
u^{+\gamma} \delta\tilde l^\beta_\gamma u^-_\beta +
x^{+\dot p}\delta c_{\dot p}^\beta u^-_\beta) x^{+\dot\alpha} -
u^{+\gamma} \delta b^{\dot\alpha}_{\gamma} -
x^{+\dot s} \delta \tilde r^{\dot\alpha}_{\dot s}
$$
\be
\delta u^+_\alpha = (u^{+\gamma}\delta \tilde l^\beta_\gamma u^+_\beta
+ x^{+\dot p} \delta c^\beta_{\dot p} u^+_\beta ) u^-_\alpha, \label{k}
\ee
and indeed
\be
\delta u^-_\alpha = 0. \label{tr}
\ee
One recognizes in (\ref{k}), (\ref{tr}) transformations of coordinates
and harmonics obtained already in section 2.2.
For the accompanying compensating transformations from the parabolic
group we get:
\be
\Delta z^{--} = u^{-\gamma}\delta \tilde l^\beta_\gamma u^-_\beta \quad
\Delta a = \textstyle -\frac{1}{2} x^{+\dot s} \delta c_{\dot s}^\beta
u^-_\beta - u^{+\gamma} \delta\tilde l^\beta_\gamma u^-_\beta ,
\ee
\be
\Delta \tilde r^{\dot\beta}_{\dot\alpha}=
-\delta \tilde r^{\dot\beta}_{\dot\alpha} +
\delta c_{\dot\alpha}^\beta u^-_\beta  x^{+\dot\beta}
\textstyle -\frac{1}{2} x^{+\dot s} \delta c_{\dot s}^\beta u^-_\beta
\delta_{\dot\alpha}^{\dot\beta} ,
\ee
\be
\Delta d = -\delta d
\textstyle -\frac{1}{2} x^{+\dot s} \delta c_{\dot s}^\beta u^-_\beta,
\ee
\be
\Delta c_{\dot\alpha}^s  = -\delta c_{\dot\alpha}^s,
\ee
and finally
\be
\Delta a^{-\dot s}= -\delta b^{\beta \dot s} u^-_\beta
+ u^{-\gamma}\delta\tilde l^\beta_\gamma u^-_\beta x^{+\dot s} \label{comp}
\ee
Of great importance is that all these transformations and manipulations
are consistent with the combined conjugation discussed in section 2.2, which
is realized on harmonics and coordinates by (\ref{ccon}) and
(\ref{cconx}).

In this form we may easily identify three complex coordinates for
our coset. The first two are $x^{+\dot\alpha}$ and the third, $z$,
may be obtained by setting
\be
u_\alpha^- = ( 1, 0), \quad  u_\alpha^+ = ( z, 1)
\ee
The transformation law for $z$ follows from eq. (\ref{k})

An important lesson is that the {\it complex} (in the common sense)
manifold is {\it real} with respect to the combined conjugation. All
transformations  must be consistent with this fact. In particular,
this compatibility condition picks out the
$Spin(5,1)$ subgroup of $Spin(6,C)$.

{\it Remark}. It is rather easy to find the
finite transformations in the normal form. For instance, conformal
transformations are written
\be
\tilde x^{+\dot\alpha} = {x^{+\dot\alpha} \over
{1 - x^{+\dot\beta} c^\sigma_{\dot\beta} u^-_\sigma }},  \quad
\tilde u^+_\alpha = u^+_\alpha +
{x^{+\dot\beta} c^\sigma_{\dot\beta} u^+_\sigma \over
{1 - x^{+\dot\beta} c^\sigma_{\dot\beta} u^-_\sigma }} u^-_\alpha  \quad
\tilde u^-_\alpha = u^-_\alpha.
\ee
They are singular at some finite value of the conformal parameter
because of gauge fixing with only one set of coordinates (i.e.
one chart) for the whole manifold\footnote{Authors are indebted
to A.~Galperin, who stimulated this comment}. As was mentioned above,
the global description of this coset can be achieved by
using 32 harmonics (instead of these 6 coordinates) defined
modulo the parabolic group transformations and obeying the unimodularity
constraint.
\subsection{Conclusions.}

An enlargement of space variables through the addition of some
harmonic (or twistorial) variables is known to admit a new kind
of analyticity, the harmonic (or twistorial) analyticity. We have
observed in the present paper that it is a facet of a quaternionic
analyticity and that there are several ways to define it. In the 4D
self-dual Yang-Mills and Einstein theories, an analyticity of this kind
replaces the standard complex analyticity of the 2D conformal theories.

The self-dual equations are conformally invariant. A remarkable
consequence of harmonic analyticity is that the $4D$ conformal group
has to be realized on the coset $CP^3$ of the {\it complexification},
$SL(4,C)$.
The reasonings are quite general: in any coset of the real group it is
impossible to have the $x$-dependent transformations of
$u^-$ and $u^+$ simultaneously analytic.
It has to be emphasized, however, that we
deal only with the ``Euclidean" conformal group $Spin(5,1)$. As a
consequence, a combined conjugation can be defined (instead of a complex
one), in the framework of which $CP^3$ is a real manifold.
It is worth mentioning that the same
phenomenon of complexification appears also in the $N=2$ and $N=3$
supersymmetric theories.

These topics will be discussed elsewhere, as well
as a more complete analysis and a classification of ``analytic" symmetries
of the self-dual equations. It is performed most effectively in the
normal gauge, taking the transformations in the normal form.
The consideration in parallel of the symmetries of
the self-dual equations and of those of the lower-dimensional integrable
systems seems to be attractive and could elucidate many subtleties.
We postpone also to future publications an investigation of the
harmonic analytic features of the self-dual gauge theories in signature
(2,2) that are expected to have intriguing
peculiarities due to a different
``non-compactness" of the corresponding rotation and conformal groups.

Note finally that $\frac {SU(2)\times SU(2)} {U(1)}$
harmonics turn out to underlie Fueter analyticity.

\subsection*{ACKNOWLEDGEMENTS}

We would like to thank sincerely C.~Devchand, A.~Galperin, P.~S.~Howe,
E.~Ivanov, L.~Michel, P.~van Nieuwenhuizen, O.~Ogievetsky, M.~Saveliev and
C.~ Vafa for valuable discussions. One of us (VO) appreciates
very much the cordial
hospitality of the Rockefeller University, where this work was begun,
and of CERN and Geneva University where it was completed. ME would also
like to thank CERN and JINR, Dubna, for hospitality.

\section*{Appendix A. Quaternionic structures}
\newcount\eqnumber
\def\cleareqnumber{\eqnumber=0}\cleareqnumber
\def\numbereq{\global\advance\eqnumber by 1 \eqno\hbox{(A--\the\eqnumber)}}
\def\name#1{\xdef#1{\the\eqnumber}}
An almost quaternionic structure is a set of three tensors of type (1,1),
${J_a}^n_m$, acting on the tangent bundle of a manifold, that represent
a basis of the quaternionic algebra (\ref{3}):

$$
J_a J_b = -\delta_{ab} +\epsilon_{abc} J_c      \numbereq
$$
Because of the non-commutativity of quaternions one has to
distinguish between {\it
left}
($L_a$) and {\it right} ($R_a$) quaternionic structures. We saw above
that the right quaternionic structures
form a two-parameter family (\ref{9}):
$$
{R_a}_{\alpha \dot\alpha}^{\beta \dot\beta} = -i u^i_{\dot\alpha}
{\sigma_a}^j_i u^{\dot\beta}_j  \delta_\alpha^\beta .   \numbereq
$$
An analogous statement is valid
with respect to the left quaternionic structures $L_a$.

An interesting observation: Given two mutually commuting
quaternionic structures, e.g. $L_a$ and $R_a$, one can construct a
one-parameter family of
quaternionic structures that interpolates between them. To do this we
construct the operator
$$
\eta=\frac{1}{2}(1-L_aR_a).     \numbereq
$$
It has the properties (that follow from quaternionic algebras for $L_a$,
$R_a$ and because they commute mutually)
$$
\eta^2=1, \hskip10mm \eta L_a \eta =R_a, \hskip10mm \eta R_a \eta=L_a.
\numbereq
$$
Now it becomes obvious that the quaternionic algebra (\ref{3}) will be
satisfied with a ``mixed'' quaternionic structure
$$
J_a= e^{c \eta} L_a e^{-c\eta}=L_a \cosh^2 c - R_a \sinh^2 c +
\epsilon_{abc} R_b L_c \cosh c \sinh c  \numbereq
$$
and commuting with this quaternionic structure
$$
J_a '= e^{c \eta} R_a e^{-c\eta}=R_a \cosh^2 c - L_a \sinh^2 c -
\epsilon_{abc} R_b L_c \cosh c \sinh c,         \numbereq
$$
where $c$ is a (real) parameter. Therefore, in 4D
space there is a 5-parameter system of quaternionic structures (2
parameters in the choice of $L_a$, 2 in that of $R_a$, and
the parameter $c$).

\section*{Appendix B. Compact cosets of non-compact groups}
\cleareqnumber
\def\numbereq{\global\advance\eqnumber by 1 \eqno\hbox{(B--\the\eqnumber)}}
Here, some mathematical definitions and statements are presented in a form
convenient for us, together with illustrations drawn from
the paper.

The Iwasawa decomposition for a non-compact semi-simple group $G$ is
(see textbooks \cite{vil}, \cite{bas})
$$
G =K A N.      \numbereq\name{\iw}
$$
Here $K, A$ and $N$ are subgroups of $G$ having the following meaning:
$K$ is the maximal compact subgroup of $G$. Denote generators of $G$ by
$\varrho$ and those of $K$ by $\kappa$.
Let $\upsilon$ be remaining generators of $G$ and $\alpha =
\{\alpha_1,\cdots,\alpha_n\}$ is a maximal Abelian subalgebra in
$\upsilon$. $A=e^\alpha$ is a commutative subgroup of $G$ generated by
$\alpha$. Finally, all generators of $G$ are decomposed in a direct
sum of eigenspaces under an adjoint action of $\alpha$,
$$[\alpha_k, \varrho_\gamma] = \gamma(\alpha_k) \varrho_\gamma,$$
$$
\varrho = \sum_{\gamma} \varrho_{\gamma}, \quad
\gamma = \{\gamma(\alpha_1), \cdots, \gamma(\alpha_n)\}.        \numbereq
$$
It is said that $\gamma$ is positive, $\gamma > 0$, if its
first non-vanishing component is positive. A space $n = \{n_{\gamma}\}$
of generators $\varrho_\gamma$ with positive $\gamma $ is a
maximal nilpotent subalgebra of $\varrho$.  $N=e^n$ is a
corresponding maximal nilpotent subgroup of $G$.

Now, the maximal solvable subgroup of $G$ is the product $  A N $.
The Borel parabolic subgroup of $G$ is
$$
B = M A N,      \numbereq
$$
where $M$ is the centralizer of the subgroup $A$ in $K$, i.e. a subgroup
of $K$ commuting with $A$.

The parabolic subgroups $P$ of $G$ are defined as those that
contain the Borel one as their subgroup. In other words,
$$
P = L A N,     \numbereq\name{\parab}
$$
where $L$ is a subgroup of the maximal compact subgroup $K$ above,
containing in turn $M$  as a subgroup. The Borel subgroup is a minimal
parabolic subgroup. It is a ``gist" of non-compactness, as can be seen through
the remarkable Borel theorem \cite{hum}:

A coset of a non-compact group $G$ modulo any of
its parabolic subgroups, $P$, is a compact space.

Moreover, parabolic subgroups $P$ can be defined as just
those such that the cosets $G \over P$ are compact. The Borel subgroup
is the smallest parabolic group.

An intuitive demonstration of this theorem is quite transparent:
$$
{G\over P} = {K A N\over L A N} = {K \over L},      \numbereq\name{\proof}
$$
$K$ and $L$ being compact. Despite being oversimplified,
this consideration is effective in
that it shows explicitly which compact manifold has been derived.

We shall now give some examples from the paper:

1. $G=SL(2,C) = \left (\begin{array}{cc} \alpha & \beta \cr
\gamma & \delta \end{array} \right), \quad
\alpha \delta - \beta \gamma = 1$ (see section 2.2)

$$ K = SU(2) =\left ( \begin{array}{cc} a & b \cr
 -\bar b & \bar a \end{array} \right), \quad  |a|^2 + |b|^2 = 1,
 $$

$$\alpha = \left (\begin{array}{cc} -\phi & 0 \cr
  0 & \phi \end{array} \right), \quad  A = e^{\alpha},
  $$

$$ n = \left (\begin{array}{cc} 0 & 0 \cr
  z & 0 \end{array} \right), \quad
[\alpha, n] = + \phi n, \quad  N =e^n. $$

The parabolic group used is
$$
P = U(1)\times A N,     \numbereq
$$
where $U(1)$ is a subgroup of $K$ with
$a=e^{-i \phi}, b=0$ .
We see that $${G\over P} = {SU(2)\over U(1)} = S^2.$$

2. $ G = Spin(5,1)$, represented by matrix (\ref{O5})
$$
\left ( \begin{array}{cc} l^\alpha_\beta  & b^{\dot\beta}_\alpha \cr
c^\beta_{\dot\alpha} & r^{\dot\beta}_{\dot\alpha} \end{array} \right )
\numbereq \name{\tens}
$$
Its maximal compact subgroup $K = Spin(5)$ is given by the same matrix
with
identification $ b_\alpha^{\dot\alpha} = \epsilon^{\dot\alpha \dot\beta}
\epsilon_{\alpha \beta} c^\beta_{\dot\beta}$ and with unimodular
$l^\alpha_\beta$ and $r^{\dot\alpha}_{\dot\beta}$.

For this case $A = \exp {\left (\begin{array}{cc} - I & 0 \cr
  0 & I \end{array} \right)}$ is the dilatation group.

Finally, $n = \left (\begin{array}{cc} 0 & 0 \cr c^\alpha_{\dot\alpha} &
O \end{array} \right ) $ are conformal boosts; $N=e^n$.

The parabolic subgroup $ P = Spin(3)\times SO(2)\times A N $ leads to a
six-dimensional coset
$$
{Spin(5,1)\over P} = {Spin(5)\over Spin(3)\times SO(2)} =
{SO(5)\over SO(3)\times SO(2)}                      \numbereq
$$
With this coset, however, conformal transformations would be
non-analytic, as explained in section 6.

3. $Spin(6,C)\sim SL(4,C)$. It is convenient to represent it again by a matrix
(B--\tens), however now with  complexified entries.

Now the maximal compact group is $K = Spin(6)\; (\sim SU(4)) $ given
by the unitarized matrix (B--\tens).

The group $A=e^{a_i \alpha_i}$ has three generators,
$$ \alpha =
\left (\begin{array}{cc} - \sigma_3  & 0 \cr   0 & 0 \end{array} \right),
\left (\begin{array}{cc}  0 & 0 \cr  0 & -\sigma_3 \end{array} \right),
\left (\begin{array}{cc} - I & 0 \cr   0 & I \end{array} \right) $$
Indices $\gamma = \{ \gamma(\alpha_1), \gamma(\alpha_2),
\gamma(\alpha_3)\} $ can be shown in a matrix form

$$\left (\begin{array}{cccc} 0\>0\>0 & 0\>\hbox{$-2$}\>0 & --+ & --- \cr
0\>2\>0 & 0\>0\>0 & -++ & -+- \cr ++- & +-- & 0\>0\>0 & 0\>0\>\hbox{$-2$} \cr
+++ & +-+ & 0\>0\>2 & 0\>0\>0 \end{array} \right) $$
where triples of indices in each entry are its indices $\gamma$.

So, there are six {\it complex} (equivalent to twelve real) generators
$n$ with positive indices. They are arranged below the main diagonal.
According to our general rule the maximal nilpotent group is $N = e^n$
and $B = A N$.

For a parabolic subgroup
$$
         P = SU(3)\times U(1)\times A N ,       \numbereq
$$
one gets a six-dimensional coset
$$
{SL(4,C)\over P} ={SU(4)\times A N \over SU(3)\times U(1)\times A N} =
{SU(4)\over SU(3)\times U(1)},
\numbereq\name{\cp3}
$$
i.e. just $CP^3$ projective space.


\begin{thebibliography}{99}
\bibitem{oldgu} F.G\"ursey, Nuovo Cim. 3 (1956) 988.
\bibitem{gur1} F.~G\"ursey and H.~C.~Tze, Ann. of Phys. (N.Y.)
128 (1980).
\bibitem{sud} A.~Sudbery, Math. Proc. Camb. Phil. Soc. 85 (1979) 199.
\bibitem{fuet} R.~Fueter, Comment. Math. Helv. 7 (1935) 307, ibid. 8
(1936) 371.
\bibitem{gurs} F.~G\"ursey, Conformal and quasi-conformal structures
in space-time, Yale prep. YCTP - P34 - 91.
\bibitem{gj} F.~G\"ursey and W.~X.~Jiang, J. Math. Phys. 33 (1992) 682.
\bibitem{gib} G.~W.~Gibbons and S.~W.~Hawking, Phys. Lett.
B 78 (1991) 430.
\bibitem{alv} L.~Alvarez-Gaum\'e and D.~Z.~Freedman, Commun. Math. Phys.
80
(1981) 443.
\bibitem{bag} J.~Bagger and E.~Witten, Nucl. Phys. B 222 (1983) 1.
\bibitem{gal} A.~Galperin, E.~Ivanov, S.~Kalitzin, V.~Ogievetsky and
E.~Sokatchev, Class. Quantum Grav. 1 (1984) 469.
\bibitem{pen} R.~Penrose, Gen.~Rel.~Grav. 7 (1976) 31.
\bibitem{ward} R.~S.~Ward, Phys. Lett. 61A (1977) 81.
\bibitem{wardw} R.~S.~Ward and  R.~O.~Wells, Twistor Geometry and Field
Theory (Cambridge Univ. Press, Cambridge, 1990).
\bibitem{penr} R.~Penrose, in ``Twistors in Mathematics and Physics", ed.
by T.~H.~Bailey and R.~J.~Baston (Cambr.Univ.Press, Cambridge, 1990,
p.1-29).
\bibitem{bas} R.~J.~Baston and M.~G.~Eastwood (The Penrose transform. Its
Interaction with Presentation Theory, Clarendon Press, Oxford, 1989).
\bibitem{galp} A.~Galperin, E.~Ivanov, V.~Ogievetsky and E.~Sokatchev,
prepr. JINR E2-85-363, (1985),  in Quantum Field Theory and
Quantum Statistics, vol.2, 233-248 (A.~Hilger, Bristol,1987)
and Ann.Phys (N.Y.) 185 (1988) 1 and 22.
\bibitem{nhi} N.~J.~Hitchin, A.~Karlhede, U.~Lindstr\"om and M.~Rocek,
Commun.Math.Phys. 108 (1987) 535.
\bibitem{og} O.~Ogievetsky, in Proc. Conf. on Group Theor. Methods in
Physics, Varna, 1987, (Berlin, Springer 1988), p.548 and Thesis,
Physical Lebedev Institute, Moscow, 1989, p.1-115.
\bibitem{sok} S.~Kalitzin and E.~Sokatchev, Class. Quantum Grav.
4 (1987) L173.
\bibitem{baggal} J.~Bagger, A.~Galperin, E,~Ivanov and V.~Ogievetsky,
Nucl. Phys. B 303 (1988) 522.
\bibitem{sin} A.~P.~Hodges, R.~Penrose and M.~A.~Singer, Phys. Lett.
B 216 (1989) 48.
\bibitem{vafa} H.~Ooguri and C.~Vafa, Mod. Phys. Lett. A 5, (1990) 1389.
\bibitem{bel} A.~Belavin and V.~Zakharov, Phys. Lett. B 73 (1978) 53.
\bibitem{cor} E.~F.~Corrigan, D.~B.~Fairlie, R.~C.~Yates and P.~Goddard,
Comm. Math. Phys. 58 (1978) 223.
\bibitem{lez} A.~Leznov and M.~Saveliev, Comm. Math. Phys. 74 (1980) 111.
\bibitem{wardint} R.~S.~Ward, Phil. Trans. Roy. Lond. Soc. A 315 (1985)
451 and in the same book as ref.[14], p.246-259.
\bibitem{hit} N.~J.~Hitchin, Proc. Lond. Math. Soc. 55 (1987) 59.
\bibitem{mas} L.~Mason and G.~A.~J.~Sparling, Phys. Lett. B 137 (1989)
29; J. Geom. and Phys. 8 (1992) 243.
\bibitem{new} S.~Chakravarti, M.~J.~Ablovitz and P.~A.~Clarkson,
Phys. Rev. Lett. 65 (1990) 1085; S.~Chakravarti, S.~Kent and
E.~T.~Newman, J. Math. Phys. 33 (1992) 382.
\bibitem{park} Q-Han Park, Phys. Lett. B 236 (1990) 429; 257 (1991) 105.
\bibitem{bak} L.~Bakas, D.~A.~Depireux, Mod. Phys. Lett. A 6 (1991) 399;
2351.
\bibitem{chau} L.- L.~Chau, I.~Yamasaka, Phys. Rev. Lett. 68 (1982) 1807.
\bibitem{tak} K.~Takasaki, W Algebra, Twistor, and Nonlinear Integrable
Systems, Kyoto prepr. KUCP-0049/92, June 1992.
\bibitem{vil} N.~Ya.~Vilenkin, A.~U.~Klimyk, Lie group representations
and special functions, In Modern Problems of Math., Fundamental Trends,
vol. 59 (VINITI, Moscow,1990) p 145-268.
\bibitem{hum} A.~Borel, Linear Algebraic Groups,(Springer, New York,
1991) New York.
\bibitem{dev} Ch.~Devchand, D.~Khetselius and V.~Ogievetsky, Heavenly
equations in harmonic space, in preparation.
\end{thebibliography}
\end{document}